# 3D hinge transport in acoustic higher-order topological insulators

Qiang Wei[1,2*], Xuewei Zhang[1,2*], Weiyin Deng[3], Jiuyang Lu[3], Xueqin Huang[3], Mou Yan[3],

Gang Chen[1,2,4†], Zhengyou Liu[5,6†], Suotang Jia[1,2]

[1]State Key Laboratory of Quantum Optics and Quantum Optics Devices, Institute of Laser spectroscopy, Shanxi University, Taiyuan 030006, China

[2]Collaborative Innovation Center of Extreme Optics, Shanxi University, Taiyuan, Shanxi 030006, China

[3]School of Physics and Optoelectronics, South China University of Technology, Guangzhou, Guangdong 510640, China

[4]Collaborative Innovation Center of Light Manipulations and Applications, Shandong Normal University, Jinan 250358, China

[5]Key Laboratory of Artificial Micro- and Nanostructures of Ministry of Education and School of Physics and Technology, Wuhan University, Wuhan 430072, China

[6]Institute for Advanced Studies, Wuhan University, Wuhan 430072, China

*These authors contributed equally to this work.
†Corresponding author. Email: chengang971@163.com; zyliu@whu.edu.cn

**The discovery of topologically protected boundary states in topological insulators opens a new avenue toward exploring novel transport phenomena. The one-way feature of boundary states against disorders and impurities prospects great potential in applications of electronic and classical wave devices. Particularly, for the 3D higher-order topological insulators, it can host hinge states, which allow the energy to transport along the hinge channels. However, the hinge states have only been observed along a single hinge, and a natural question arises: whether the hinge states can exist simultaneously on all the three independent directions of one sample? Here we theoretically predict and experimentally observe the hinge states on three different directions of a higher-order topological phononic crystal, and demonstrate their robust one-way transport from hinge to hinge. Therefore, 3D topological hinge transport is successfully achieved. The novel sound transport may serve as the basis for acoustic devices of unconventional functions.**

Transport is one of the most fundamental and important notions in condensed-matter physics and material science, and it lays the foundations for almost all devices in application [1-4]. The emergence of topological insulators (TIs) provides an unprecedented opportunity for discovering the state-of-the-art transport behaviors of electrons [5-7] or waves [8-15], advancing the conventionals, thanks to the topological-protected boundary states they host. The boundary states, which can survive the disorders and impurities, exhibit the robust one-way feature [6, 7]. This shows great promise in low-loss devices of electronic and classical wave systems, such as thermoelectric converter [16, 17], topological negative refraction [18, 19], and topological laser [20].

Of special interest recently is the higher-order TIs, which respect a generalized bulk-boundary correspondence. It states that the $d$ D $n$ th-order TIs possess the boundary states of $(d-n)$ D [21-24]. In particular, for the 3D second-order TI, there exist the topological-protected hinge states [25], which localize on the hinges of the system. To date, the hinge states have only been observed along single hinges [26, 27]. It raises a natural question: whether the hinge states can exist simultaneously along all the three independent directions in the higher-order TIs? Recently, such hinge states have been proposed theoretically [28-32], but it is great challenge to experimentally observe these states and associated 3D one-way transport in electronic systems [33].

In this work, we theoretically predict and experimentally observe the helical hinge states along three independent directions in a higher-order topological phononic crystal (PC). Specifically, it is constructed by stacking a bilayer hexagonal lattice, with opposite on-site energies for the two layers and different intracell and intercell layer couplings. By interchanging the on-site energies of the bilayer or the intracell and intercell layer couplings, four different phases are obtained, which possess gapped surface and interface states. The Jackiw-Rebbi mechanism [34] results in the hinge states along all the three independent directions, as desired. More importantly, based on these independent hinge states, we demonstrate the 3D robust one-way transport from hinge to hinge. The theoretical, simulated, and experimental results are in good agreement. Our work not only unveil the exotic topological states of matter, but also opens an avenue in the design of new devices.

To begin with, we introduce a tight-binding model by A-A stacking of a bilayer hexagonal lattice. As shown by the left panel in Fig. 1a, a unit cell hosts four sites $A_i$ (magenta spheres) and $B_i$ (purple spheres), where $A, B$ denote the sublattices of each layer and $i = 1,2$ is the layer degree of freedom. The sites $A_1$ and $B_2$ have the same on-site energy $m_0$, while the other sites have the opposite value. At each layer, the nearest-neighbor hoppings, labeled by $t_0$ (gray tubes), occur for both the sublattices A and B, while for two adjacent layers, the hoppings, denoted respectively by $t_1$ (yellow tubes) and $t_2$ (cyan tubes), emerge only for the sublattice A. In the basis of $(A_1, B_1, A_2, B_2)$, the Bloch Hamiltonian

$$\mathcal{H}(\boldsymbol{k}) = \begin{pmatrix} m_0 & f & g & 0 \\ f^* & -m_0 & 0 & 0 \\ g^* & 0 & -m_0 & f \\ 0 & 0 & f^* & m_0 \end{pmatrix}, \quad (1)$$

with $f = t_0[1 + 2\exp(-i\sqrt{3}k_y a/2)\cos(k_x a/2)]$ and $g = t_1 + t_2 \exp(ik_z h)$, where $\boldsymbol{k} = (k_x, k_y, k_z)$ is the Bloch wavevector, and $a$ and $h$ are the lattice constants in the $x$-$y$ plane and along the $z$ direction, respectively. The right panel in Fig. 1a shows the first Brillouin zone (BZ) and its projected surface BZs in the $k_x$-$k_y$ and $k_x$-$k_z$ planes.

For the Hamiltonian in equation (1), the time-reversal symmetry is conserved, while the sublattice symmetry and the mirror symmetry along the $z$ direction are broken. As a result, the bulk band at the point H (or H′) opens a gap, $\sqrt{m_0^2 + (t_1 - t_2)^2} - |m_0|$, between the first and second bands, as shown in Fig. 1b. This gap closes for $t_1 = t_2$. In the case of $t_1 < t_2$, this Hamiltonian has gapped surface states in the $k_x$-$k_y$ plane (red lines in Fig. 1c), which have the energy extreme at the valley $\bar{K}$ (or $\bar{K}'$). These surface states are located respectively on the top and bottom surfaces and share identical dispersions. For the top one, they are governed, near the valley $\bar{K}$, by an effective Hamiltonian $\mathcal{H}_s = v_s(\bar{q}_x \sigma_x + \bar{q}_y \sigma_y) + m_s \sigma_z$, where $v_s = (\sqrt{3}t_0/2)\sqrt{1 - (t_1/t_2)^2}$, $m_s = m_0$ is the effective mass, $\bar{q}_x$ and $\bar{q}_y$ are the infinitesimal momenta, and $\sigma_i$ $(i = x, y, z)$ is the Pauli matrix acting on the sublattice degree of freedom; see Supplementary Section I for details. This effective

Hamiltonian supports a valley Chern number $C_{\bar{K}} = \text{sgn}(m_s)/2$. We emphasize that the effective mass in the $k_x$-$k_y$ surface can be well tuned by controlling the on-site energies of the sublattices. This controllable effective mass can help us construct the hinge states, as will be shown below.

We introduce two phases I and II with the opposite effective masses, i.e., $\pm m_s$. When these two phases touch at the $x$-$z$ plane (left panel of Fig. 1d), the hinge states along the $x$ direction in the top surface (blue arrows) are formed by the Jackiw-Rebbi mechanism [33]. These hinge states are protected by an integer topological invariant $\Delta C_{\bar{K}} = C_{\bar{K}}^{\text{I}} - C_{\bar{K}}^{\text{II}} = \text{sgn}(m_s)$ [11], and are dominated by an effective Hamiltonian $\mathcal{H}_h^x = v_s \bar{q}_x s_z$, where $s_z$ is the Pauli matrix acting on the valley degree of freedom; see Supplementary Section II for details. This effective Hamiltonian demonstrates the helical property of the hinge states, which can also be verified from their dispersions in Supplementary Fig. S2. The hinge states in the bottom surface (green arrows) are the same as those in the top surface, but with an opposite topological invariant. Similarly, the hinge states along the $y$ direction can also be produced when these two phases touch at the $y$-$z$ plane.

Note that when the phases I and II touch, there also exists gapped interface states, which have the energy extreme at the valley $\bar{H}$ (or $\bar{H}'$), as shown by the red lines in Supplementary Fig. S3. In the vicinity of the valley $\bar{H}$, they are dominated by an effective Hamiltonian $\mathcal{H}_{\text{I-II}} = \sqrt{3} t_0 \bar{q}_x' \tau_x/2 + t_2 \bar{q}_z' \tau_y/2 + m_{\text{I-II}} \tau_z$, where $\bar{q}_x'$ and $\bar{q}_z'$ are the infinitesimal momenta, $m_{\text{I-II}} = (t_1 - t_2)/2 < 0$ is also the effective mass controlled here by the interlayer couplings, and $\tau_i$ $(i = x, y, z)$ is the Pauli matrix acting on the layer degree of freedom; see Supplementary Section III for details. In this case, a valley Chern number, $C_{\bar{H}} = -1/2$, is obtained. We further introduce two extra phases III and IV, in which the phase III (IV) has a similar property as that of the phase I (II) but with $t_1 > t_2$. When the phases III and IV touch also at the $x$-$z$ plane, similar interface states emerge and have an effective mass $m_{\text{III-IV}} = (t_1 - t_2)/2 > 0$. Since $m_{\text{I-II}} < 0$ and $m_{\text{III-IV}} > 0$, when the interfaces of I-II and III-IV touch (middle of Fig. 1d), the helical hinge states along the $z$ direction (red arrows) are also produced by the Jackiw-Rebbi mechanism [33]. They are protected by an integer topological invariant $|\Delta C_{\bar{H}}| = 1$ [11] and are governed by an effective Hamiltonian $\mathcal{H}_h^z = t_2 \bar{q}_z' s_z/2$. The corresponding dispersions are plotted in Supplementary Fig. S4.

Since the on-site energy of the phase I (II) has the same as that of the phase III (IV), no interface states of I-III and II-IV emerge. Also, for $t_1 > t_2$ in the phases III and IV, no surface states in the $k_x$-$k_y$ plane can be found. As a result, for the structure in the middle of Fig. 1d, the hinge states only along the $x$ and $z$ directions can be produced. Fortunately, by introducing a L-shaped design for the phase II based on the above structure, the hinge states in three independent directions can be achieved successfully, as shown by the right panel in Fig. 1d. Notice that all the interfaces are here designed as the zigzag type, the new hinge states are generated along the direction that has $30°$ with respect to the $y$ axis. It is worth mentioning that all the hinge states emerge around zero energy (Supplementary Figs. S2 and S4), which indicates the emergence of the 3D hinge transport.

We now experimentally verify our findings in PCs. Figure 2a shows the side (left panel) and top (right panel) views of a unit cell for the phase I. This bilayer hexagonal unit cell is formed by two triangular prism scatterers, which have the same side length ($s_0 = 12.85$ mm) and height ($h_0 = 6.00$ mm), but opposite rotation angles, $\pm 30°$, with respect to the $x$ axis. Two types of the triangular holes are drilled through the neighboring supporting plates with the side lengths and heights, $s_1 = 3.46$ mm, $h_1 = 6.50$ mm and $s_2 = 8.66$ mm, $h_2 = 10.40$ mm. The lattice constants in the $x$-$y$ plane and along the $z$ direction are $a = 17.32$ mm and $h = 28.90$ mm, respectively. The unit cell of the phase II has the same as that of the phase I but with the opposite rotation angle for each layer, while the unit cell of the phase III (IV) is the same as that of the phase I (II) but with the different side lengths $s_1 = 8.66$ mm and $s_2 = 3.46$ mm.

We first observe the hinge transport along the $x$ direction by printing a PC sample with $30 \times 15 \times 6$ unit cells shown by the left panel in Fig. 2b. The right panel reflects the structure of this PC sample by stacking the unit cells of the phases I and II along the $z$ direction. Since the PC sample has two hinge states located respectively in the top and bottom surfaces, we should separately measure the acoustic pressure fields of these surfaces. When the point source is placed at the left end (green star) to excite the hinge state in the top surface (see Methods for details), the hinge state dispersion along the $k_x$ direction is obtained by Fourier transforming the measured acoustic pressure fields, as shown by colors in Fig. 2c. The left panel in Fig. 2d presents the measured acoustic pressure fields at the excitation frequency of 10.0 kHz. It shows that the sound wave propagates to the right end along the hinge in the top surface and decays rapidly in the vertical plane. The experimental data agree well with the full-wave simulations (Methods), as shown by the right panel in Fig. 2d. If the point source is placed at the right end, the sound wave propagates to the left end at the same excitation frequency (Supplementary Fig. S5), which, together with Fig. 2d, verifies the helical property of the hinge states. The hinge states in the bottom surface are similar to those in the top surface, as shown in Figs. 2e and 2f.

We then observe the hinge transport along the $z$ direction by printing another PC sample with $20 \times 16 \times 10$ unit cells shown in Fig. 3a. The corresponding structure is illustrated in Fig. 3b. This PC sample is constituted by four phases I, II, III, and IV. When the point source is placed at the intersection point of the four phases in the top surface, the measured hinge state dispersion along the $k_z$ direction is shown by colors in Fig. 3c. Due to the helical property, the hinge state can also be excited by the point source in the bottom surface and the corresponding dispersion is measured in Fig. 3d. The observed dispersions match well with the simulation ones. In Fig. 3e, we observe (left panel) and simulate (right panel) the acoustic pressure fields at the excitation frequency of 10.0 kHz, showing that the sound wave propagates from the bottom to the top along the hinge of the $z$ direction.

Finally, we observe 3D hinge transport by printing a PC sample with $24 \times 16 \times 4$ unit cells shown in Fig. 4a. By placing the point source at the bottom surface, we measure the acoustic pressure field distributions in the top and bottom surfaces as well as the interfaces of I-III and II-IV. The left panel of Fig. 4b presents the experimental data at the excitation frequency of 10.0 kHz. It shows that the sound wave first propagates along the hinge of the bottom $x$-$y$ plane, then turns into the hinge of the $z$ direction, and finally travels along the hinge of the top $x$-$y$ plane (gray-arrowed lines). This experimental observation, matching with the simulation shown by the right panel of Fig. 4b, demonstrates definitely the 3D hinge transport, as expected.

In Fig. 4c, we measure the transmissions for two cases of the weak disorders to demonstrate the robustness of the 3D hinge transport. At this time, the point source and the receiver are placed at the incident port on the bottom surface and outgoing port on the top surface, respectively. The red curve shows the experimental data when the disorder is generated by rotating randomly the scatterers at the range of $\pm 10°$ in the bottom surface (left panel in Fig. 4d), while the blue curve represents experimental data when the disorder induced by a similar way is in the third and fourth layers (right panel in Fig. 4d). These curves show that for weak disorders, the transmissions within the frequency range of 9.5~10.1 kHz almost remain unchanged, verifying the robustness of the 3D hinge transport against the disorders. This is a universal property caused by the intrinsic topological feature of the acoustic TIs.

In summary, we have theoretically predicted and experimentally observed the helical hinge states along three independent directions in a higher-order topological PC. According to the Jackiw-Rebbi mechanism, these hinge states are constructed by the gapped surface and interface states with the controllable effective masses. We have also demonstrated the 3D hinge transport with one-way feature. The novel sound transport may serve as the basis for acoustic devices of unconventional functions. In addition, our method can be easily extended to unveil exotic topological states of matter, such as the high-order topological semimetal with rich hinge states [35].

**Data availability**

The data that support the plots within this paper and other findings of this study are available from the corresponding author upon reasonable request.

**Acknowledgements**

This work is supported by the National Key R & D Program of China (Grant Nos. 2017YFA0304203 and 2018YFA0305800), the National Natural Science Foundation of China (Grant Nos. 11890701, 11974120, 11974005, 12034012, 12074128, and 12074232), and Shanxi "1331 Project" Key Subjects Construction.

**Author contributions**

All authors contributed extensively to the work presented in this paper.

**Competing financial interests**

The authors declare no competing financial interests.

**Methods**

**Numerical simulations.** All full-wave simulations of the PCs are performed by the commercial finite-element-method software (COMSOL Multiphysics). Since the photopolymer materials (with the mass density $1160 \text{ kg} \cdot \text{m}^{-3}$ and the sound velocity $2540 \text{ m} \cdot \text{s}^{-1}$) have a huge impedance mismatch with air (with the mass density $1.293 \text{ kg} \cdot \text{m}^{-3}$ and the sound velocity $337 \text{ m} \cdot \text{s}^{-1}$), they can be considered as hard boundaries in the simulations. In addition, the hinge state dispersions along the $k_x$ ($k_z$) direction are calculated using a ribbon structure only with the periodic boundary condition in the $x$ ($z$) direction.

**Experimental measurements.** A sub-wavelength headphone is used as a point source to excite the acoustic pressure field. This headphone has the diameter $5.6 \text{ mm}$ and is connected to a network analyzer with the model: Keysight 5061B. To measure the acoustic pressure field, a sub-wavelength microphone is manually inserted into the sample as a receiver. This microphone with the diameter $3.0 \text{ mm}$ is bonded to the tip of the stainless steel rod and is connected to the network analyzer. The network analyzer not only generates the excitation signal that a sinusoidal wave sweeps from $\sim 7.0 - 13.0 \text{ kHz}$, but also collects the recorded signals (S-parameter $S_{21}$) with average processing of 16 times. By further Fourier transforming the measured acoustic pressure fields, the hinge state dispersions along the $k_x$ and $k_z$ directions are obtained. All the contours in Figs. 2c, 2d, 2e, 2f, 3c, 3d, 3e, and 4b, 4c are normalized by their maxima.

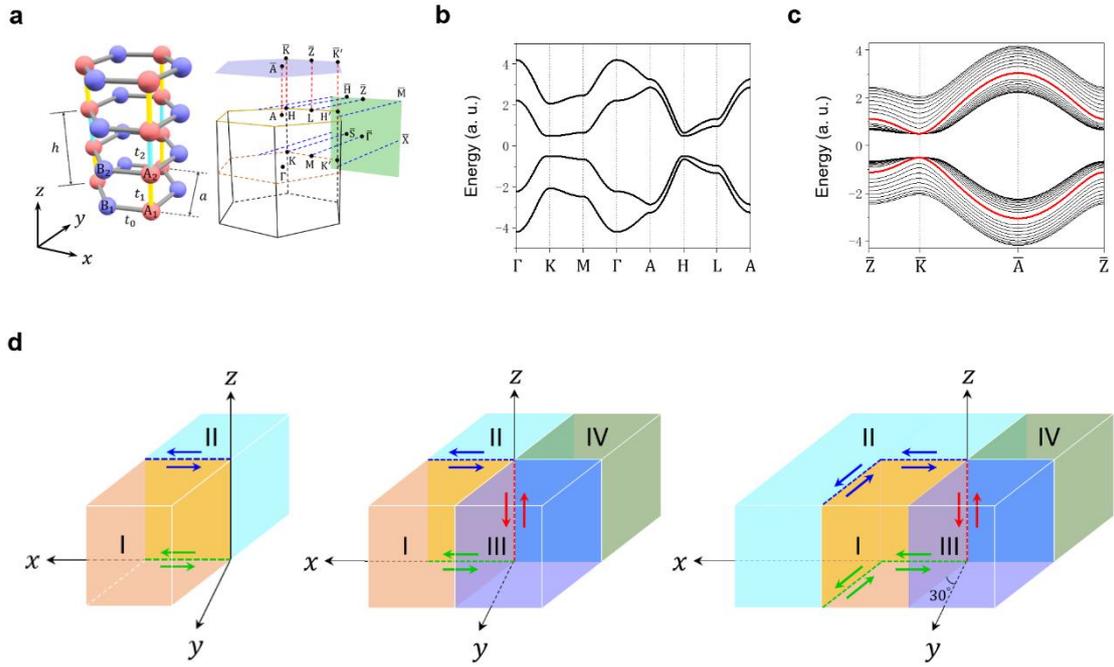

**Figure 1 | Tight-binding model of generating 3D hinge transport. a**, Left panel: schematics of a bilayer hexagonal lattice with four sites in the unit cell. Right panel: the first BZ and associated projected surface BZs. **b**, Bulk band dispersions along the high-symmetry lines. **c**, Surface state dispersions in the projected BZ in the $k_x$-$k_y$ plane. The red (black) curves represent the dispersions of the surface (bulk) states. The plotted parameters in **b** and **c** are chosen as $t_0 = -1$, $m_0 = -0.5$, $t_1 = -0.8$, and $t_2 = -1.2$ in arbitrary unit (a.u.). **d**, Schematics of the hinge states. The left panel, middle, and right panel show them along the $x$ direction in the top (blue arrows) and bottom (green arrows) surfaces, the $x$ and $z$ (red arrows) directions, and three independent directions, respectively.

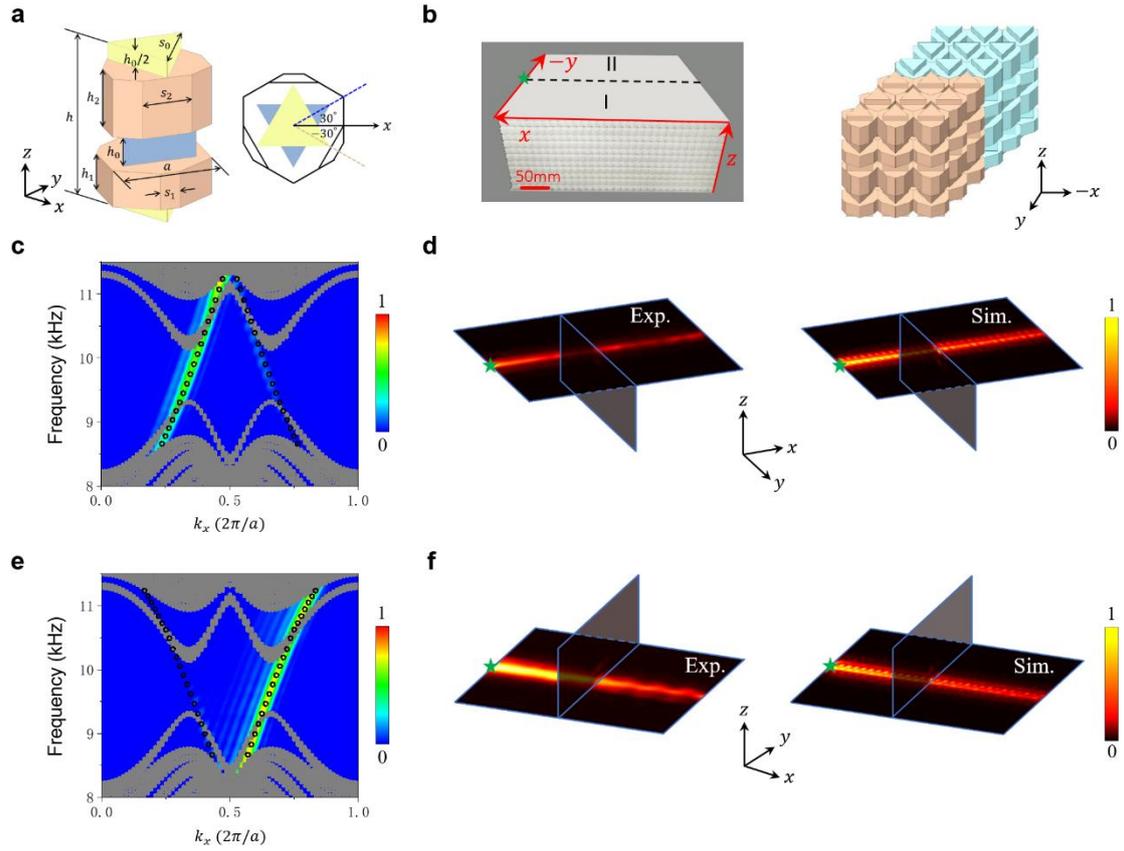

**Figure 2 | Hinge transport along the $x$ direction. a**, The side (left panel) and top (right panel) views of the unit cell for the phase I. **b**, Left panel: a photo of the 3D-printing PC sample. Right panel: the schematic structure of the PC sample. **c**, Projected band dispersions along the $k_x$ direction in the top surface. The color maps represent the experimental data, while the black (gray) circles reflect the simulated results of the hinge (bulk and surface) states. **d**, Measured (left panel) and simulated (right panel) acoustic pressure field distributions of the hinge states at the excitation frequency of 10.0 kHz. The green stars denote the positions of the point source. **e,f**, Projected band dispersions along the $k_x$ direction in the bottom surface and the corresponding acoustic pressure field distributions of the hinge states.

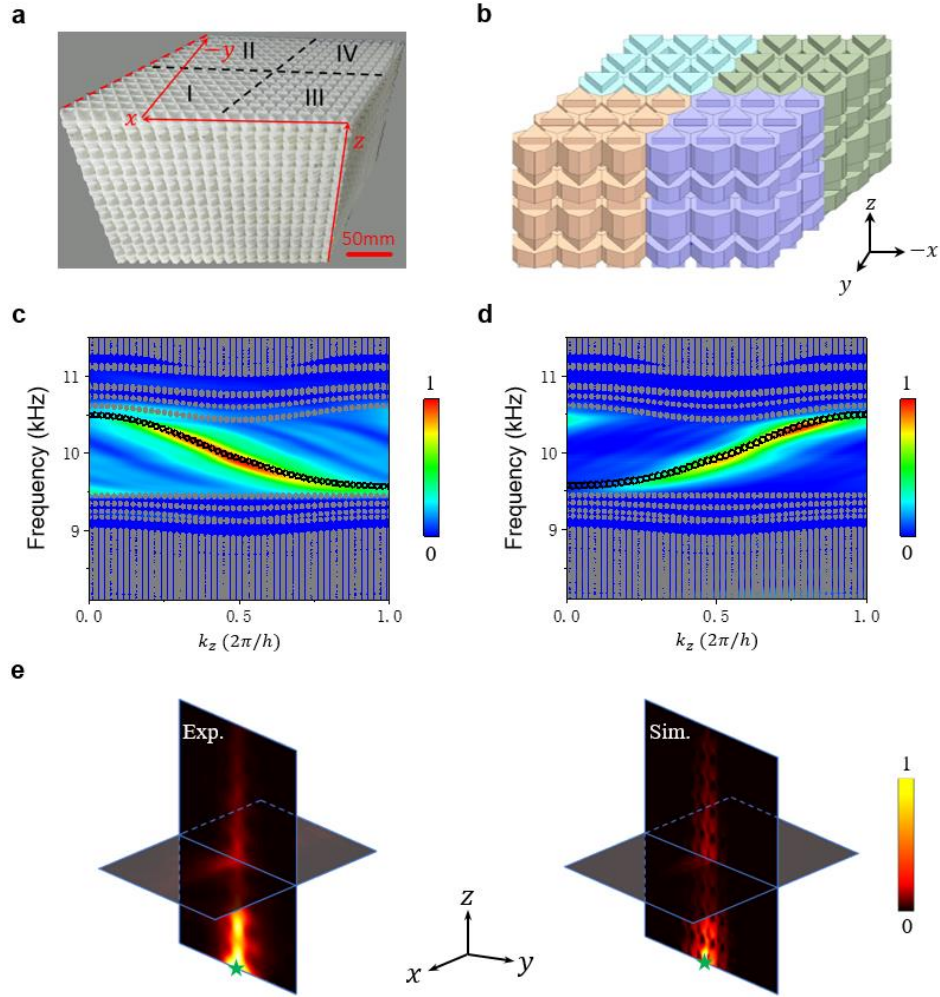

**Figure 3 | Hinge transport along the $z$ direction. a**, A photo of the 3D-printing PC sample with four phases I, II, III, and IV. **b**, Schematic structure of the PC sample. **c,d**, Projected band dispersions along the $k_z$ direction, when the point sources are placed at the intersection points of four phases in the top and bottom surfaces, respectively. The color maps represent the experimental data, while the black (gray) circles reflect the simulated results of the hinge (bulk and surface) states. **e**, Measured (left panel) and simulated (right panel) acoustic pressure field distributions of the hinge states at the excitation frequency of 10.0 kHz. The green star denotes the position of the point source.

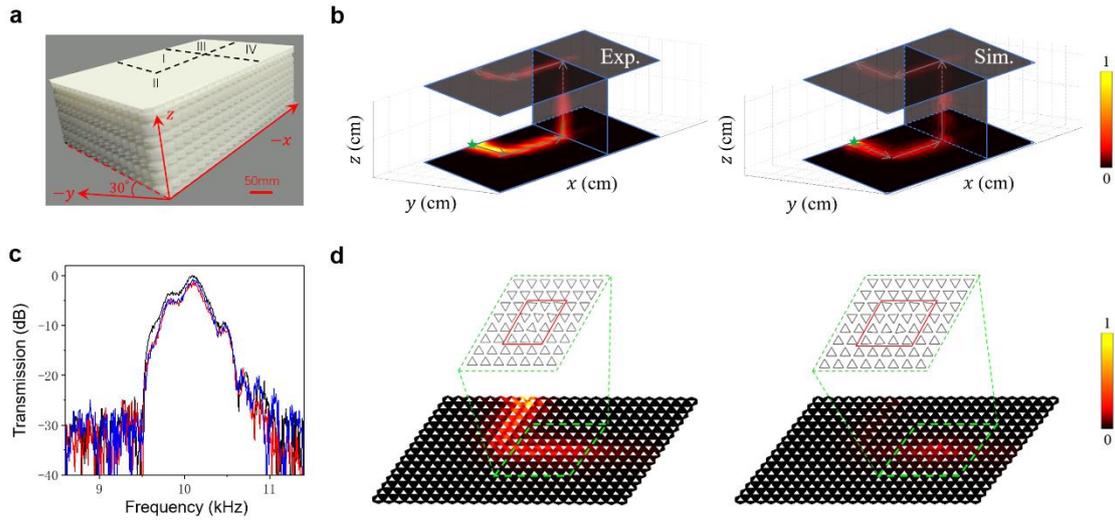

**Figure 4 | 3D hinge transport. a**, A photo of the printing PC sample with a L-shaped design for II. **b**, Measured (left panel) and simulated (right panel) acoustic pressure field distributions at the excitation frequency of 10.0 kHz. The gray-arrowed lines reflect the one-way transport of the sound wave along the hinges of the sample. The green stars denote the positions of the point source. **c**, Measured transmissions with (red and blue curves) and without (black curve) the disorders. **d**, Left and right panels: configurations of two cases of the weak disorders (red boxes), corresponding to the red and blue curves in **c**, respectively.